\documentclass[a4paper,twocolumn,showpacs]{revtex4-1}
\usepackage[latin1]{inputenc}
\usepackage{amsmath,amssymb}
\usepackage{theorem}
\usepackage{graphicx}
\usepackage{color}
\usepackage{wasysym}

\definecolor{blue}{rgb}{0,0,1}
\definecolor{green}{rgb}{0,1,0}
\definecolor{red}{rgb}{1,0,0}
\definecolor{vio}{rgb}{1,0,1}
\definecolor{ama}{rgb}{1,1,0}
\newcommand{\bc}{\begin{center}}
\newcommand{\ec}{\end{center}}
\newcommand{\be}{\nopagebreak[3]\begin{equation}}
\newcommand{\ee}{\end{equation}}
\newcommand{\ba}{\nopagebreak[3]\begin{eqnarray}}
\newcommand{\ea}{\end{eqnarray}}

\begin{document}

\title{
Gravitational lens optical scalars \\
in terms of energy-momentum distributions
}
 
\author{
Emanuel Gallo 
and
Osvaldo M. Moreschi
\\
\vspace{3mm}
\small Facultad de Matem\'atica, Astronom\'{i}a y F\'{i}sica, FaMAF, Universidad Nacional de C\'{o}rdoba\\
\small Instituto de F\'{i}sica Enrique Gaviola, IFEG, CONICET\\
\small Ciudad Universitaria, (5000) C\'{o}rdoba, Argentina. \\
\vspace{3mm}
}


\begin{abstract}
This is a general work on gravitational lensing.
We present new expressions for the optical scalars and the deflection angle
in terms of the energy-momentum tensor components of matter distributions.
Our work generalizes standard references in the literature where normally
stringent assumptions are made on the sources.
The new expressions are manifestly gauge invariant, since they are presented
in terms of curvature components.
We also present a method of approximation for solving the lens equations, that
can be applied to any order.

\end{abstract}

\maketitle


\vspace{5mm}

%
\section{Introduction}

There being so many excellent publications that had covered the study of gravitational
lensing, our justification for another general article on the subject comes from the fact 
that we present new expressions for the optical scalars and deflection angle for a 
wide variety of matter distributions in terms of the matter components.

Gravitational lensing has became a significant tool to make progress in our knowledge on the 
matter content of our Universe. 
In particular, there is a large number of works that use gravitational lensing 
techniques in order to know how much mass are in galaxies or clusters of galaxies. 
One of the most exiting results was to reaffirm the need for some kind of
dark matter, that appears to interact with the barionic matter only through gravitation.

The question in which there is yet not general agreement is on 
the very nature of this dark matter.
The most common conception is that it is based on collisionless particles\cite{Weinberg08}, and 
where the pressures are negligible.
However in the context of cosmological studies, one often recurs to models of dark matter
in terms of scalar fields\cite{Matos00,Urena02,Arbey03,Nucamendi01}. There is also the possibility that dark matter were 
described in terms of spinor fields\cite{Gu07}.

One method to study the nature of dark matter consists in 
observing the deformation of images of galaxies behind a matter distribution that is the
source of a gravitational lens.

The fact that gravitational lensing can be useful for the study of the nature of dark matter
has been emphasized many times, in particular in respect to the question of its
equation of state\cite{Faber06,Bharadwaj03,Nunez10}.

\begin{figure}[h]\label{f:fig0}
\includegraphics[clip,width=0.4\textwidth]{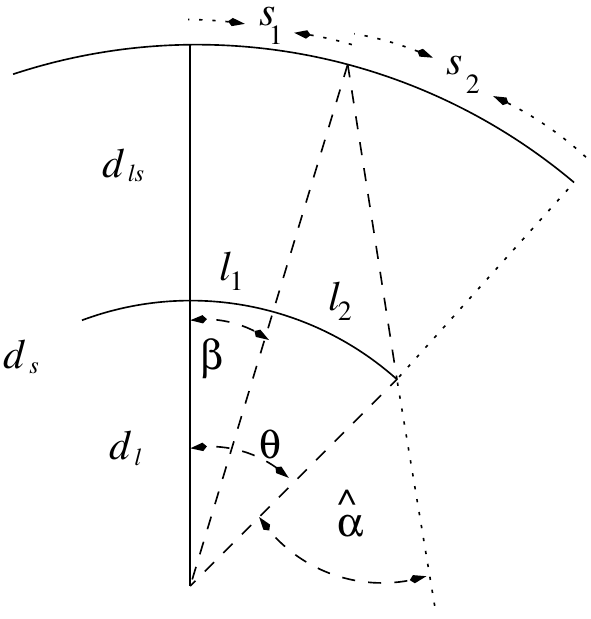}
\caption{This graph shows the basic and familiar angular variables in terms of a simple
flat background geometry.
The letter $s$ denote sources, the letter $l$ denotes lens and the observer is
assumed to be situated at the apex of the rays.}
\end{figure}

In many astrophysical situations, the gravitational effects on light rays is weak, and 
the source and observer are far away from the lens, therefore they are studied under the 
formalism of weak field and thin gravitational lenses. The basic and familiar variables in this 
discussion are shown in figure  \ref{f:fig0}.

In this framework the lens equation reads
\begin{equation}
\beta^a=\theta^a-\frac{d_{ls}}{d_s}\alpha^a.
\end{equation}

The differential of this equation can be written as
\begin{equation}\label{eq:standarlensequation}
\delta \beta^a=A^a_{\; b}\,  \delta\theta^b,
\end{equation}
where the matric $A^a_{\; b}$ is in turn expressed by
\begin{equation}
A^a_{\; b} =
\left( {\begin{array}{cc}
 1-\kappa-\gamma_1 & -\gamma_2  \\
 -\gamma_2& 1-\kappa+\gamma_1  \\
 \end{array} } \right);
\end{equation}
where the optical scalars $\kappa$, $\gamma_1$ and $\gamma_2$, are known as 
convergence $\kappa$ and shear components $\{\gamma_1,\gamma_2\}$, and have the information 
of distortion of the image of the source due to the lens effects.
  
It is somehow striking that in
most astronomical works on gravitational lensing, it is assumed that the 
lens scalars and deflection angle, can be obtained from a Newtonian-like potential function. 
These expressions although are easy to use, have some limitations:
\begin{itemize}
\item They neglect more general distribution of energy-momentum tensor $T_{ab}$, in particular 
they only take into account the timelike component of this tensor.
In this way they 
severely restrict the possible candidates to dark matter that can be studied with these 
expressions. 
\item They are not expressed in terms of gauge invariant quantities.
\item Since these expressions are written in terms of a potential function, 
it is not easily seen how  different components of $T_{ab}$ contribute in the 
generation of these images.
\end{itemize}

Moreover, most of them assume from the beginning that thin lens is a good approximation.

In other cases in which the thin lens approximation is not used\cite{Bernardeau10}, the results are
presented in a way in which gauge invariance is not obvious; however see \cite{Frittelli00a}.

In this paper we extend the work appearing in standard references on gravitational 
lensing\cite{Schneider92,Seitz94,Wambsganss98,Bartelmann10} and
present new expressions that do not suffer from 
the limitations mentioned above. In particular we present gauge invariant expressions
for the optical scalars and deflection angle for some general class of matter
distributions.
In this first work on the subject, we study weak field gravitational lensing over a flat background.

In section \ref{sec:general} we present the general setting, starting from the geodesic deviation 
equation, where we fix some of our notation, and obtain gauge invariant expressions for the lens scalars. 
We also present a method of approximation for solving the lens equations, that
can be applied to any order.
In section \ref{sec:axisymm} and  \ref{sec:thin}, we will restrict the study to axially symmetric lens, and we 
present expressions for the lens scalars and deflection angle in the thin lens case. 
In section \ref{sec:spher}, we concentrate in the spherically symmetric case, and after a 
general study of this geometry, we obtain expressions in terms of the energy-momentum 
distribution of these optical quantities.   
We end with a summary final section \ref{sec:final} and a couple of appendices.

\section{Integrated expansion and shear}\label{sec:general}
\subsection{General equations: The geodesic deviation equation}
Let us consider the general case of a null geodesic starting from the
position $p_s$ (source) and ending at $p_o$ (observer).
Let us characterize the tangent vector as $\ell= \frac{\partial}{\partial \lambda}$;
so that
\begin{equation}
 \ell^b \nabla_b \ell^a = 0 ;
\end{equation}
that is, $\lambda$ is an affine parameter.

We can now consider also a continuous set of nearby null geodesics.
This congruence of null geodesics can be constructed in the following way.
Let $S$ be a two dimensional spacelike surface (the source image)
such that the null vector $\ell$ is orthogonal to $S$.
Next we can generalize  $\ell$ to be a vector field in the vicinity of the initial geodesic
in the following way: let the function $u$ be defined so that it is
constant along the congruence of null geodesics emanating orthogonally to $S$ and
reaching the observing point  $p_o$.
Then, without loss of generality we can assume that
\begin{equation}
 \ell_a = \nabla_a u ;
\end{equation}
which implies that the congruence has zero twist.

We can complete to a set of null tetrad, so that $m^a$ and $\bar m^a$
are tangent to $S$. At other points $m^a$ is chosen so that it is tangent to the surfaces $u=$constant and
$\lambda=$constant.
Then a deviation vector at the source image can be expressed by
\begin{equation}
 \varsigma^a = \varsigma \bar m^a + \bar \varsigma m^a .
\end{equation}
In order to propagate this deviation vector along the null congruence
one requires, that its Lie derivate vanishes along the congruence;
that is
\begin{equation}
 \mathfrak{L}_\ell \varsigma^a = 0 ;
\end{equation}
which is equivalent to 
\begin{equation}\label{eq:liecovar}
 \ell^b \nabla_b \varsigma^a -  \varsigma^b \nabla_b \ell^a =0 .
\end{equation}
From this it can be proved that
\begin{equation}
 \ell( \ell_b \varsigma ^b) = 0 .
\end{equation}

The expansion and shear of the congruence are defined\cite{Pirani64}
respectively by
\begin{equation}
 \theta = \frac{1}{2} \nabla_a \ell^a 
\end{equation}
and
\begin{equation}
 |\sigma| = \sqrt{\frac{1}{2} \nabla_{(a} \ell_{b)} \nabla^a \ell^b - \theta^2 } ;
\end{equation}
whose relation to the spin coefficient\cite{Geroch73} quantities $\rho$ and $\sigma$ is
given by
\begin{equation}
 \theta = -\frac{1}{2} (\rho + \bar\rho)
\end{equation}
and
\begin{equation}
 |\sigma|^2 = \sigma \bar\sigma ;
\end{equation}
with certain abuse of the notation on the first appearance of $|\sigma|$.

Let us now calculate the covariant derivative of $\varsigma^a$ in the direction
of $\ell$,
\begin{equation}
 \ell^b \nabla_b \varsigma^a =
\ell(\varsigma) \bar m^a + \varsigma \ell^b \nabla_b \bar m^a + \text{c.c.} ;
\end{equation}
while
\begin{equation}
 \varsigma^b \nabla_b \ell^a = \varsigma \bar m^b \nabla_b \ell^a +\text{c.c.} ;
\end{equation}
where c.c. means complex conjugate.

Let us note that
\begin{equation}
 \bar m^b \nabla_b \ell^a =
(- \beta' + \bar \beta ) \ell^a
- \bar \sigma m^a
- \rho  \bar m^a
\end{equation}
and
\begin{equation}
 \ell^b \nabla_b \bar m^a =
(\bar\epsilon - \epsilon)  \bar m^a
 - \bar\kappa n^a
 - \tau'\ell^a ;
\end{equation}
where we are using the GHP\cite{Geroch73} notation for the spin coefficients.
In our case one has
\begin{equation}
 \kappa = 0 ;
\end{equation}
since $\ell$ is geodesic.
{Notice} that the Lie derivative of vector $m$ in the direction of $\ell$ is
\begin{equation}
\left[ \ell ,m\right] ^{a}=\left( \bar{\rho}+\epsilon -\bar{\epsilon}\right)
m^{a}+\sigma \bar{m}^{a}+\left( \bar{\beta}^{\prime }-\beta -\bar{\tau}%
^{\prime }\right) \ell ^{a} .
\end{equation}
Then, since the Lie transport of $m$ in the direction of $\ell$ should not have any $\ell$ component;
due to the fact that $m$ are always tangent to the surfaces $u=$constant and $\lambda=$constant;
one obtains that
\begin{equation}
 \tau'= \beta'- \bar \beta .
\end{equation}
Therefore, from eq.(\ref{eq:liecovar}) one has
\begin{equation}
\begin{split} 
&\ell(\varsigma) \bar m^a 
+\varsigma  (\bar\epsilon - \epsilon)  \bar m^a\\
&+\varsigma\left(  \bar \sigma m^a
+ \rho  \bar m^a\right)
+ \text{c.c.}=0
;
\end{split}
\end{equation}
which implies
\begin{equation}
 0= \ell(\varsigma) +\varsigma  (\bar\epsilon - \epsilon)
+ \varsigma \rho
+ \bar \varsigma \sigma .
\end{equation}

Using the GHP notation one can write the previous equation as
\begin{equation*}
 0= \text{\Thorn}(\varsigma) 
+ \varsigma \rho
+ \bar \varsigma \sigma ;
\end{equation*}
where \Thorn \, is the well behaved derivation of type $\{1,1\}$ in the direction of $\ell$.
In order to have simple relations in terms of coordinate derivatives in the direction of $\lambda$,
the complex phase of $m$ and $\bar m$ can be chosen so that $\epsilon = 0$;
so that finally one has
\begin{equation}
 \ell(\varsigma) = \frac{\partial \varsigma}{\partial \lambda} = - \varsigma \rho
- \bar \varsigma \sigma .
\end{equation}
We see then that $\rho$ determines the instantaneous expansion, and $\sigma$
determines the instantaneous shear of the congruence.

Let us recall from the GHP equations\cite{Geroch73} that
\begin{equation}
 \ell(\rho) = \rho^2 + \sigma  \bar\sigma + \Phi_{00} ,
\end{equation}
and
\begin{equation}
 \ell(\sigma) = (\rho + \bar\rho) \sigma + \Psi_{0} .
\end{equation}

Defining the matrix $P$ from
\begin{equation}
 P = 
\left(
\begin{array}{cc}
\rho & \sigma \\
\bar\sigma & \bar\rho
     \end{array}
\right) ;
\end{equation}
one has
\begin{equation}
 \ell(P) = P^2 + Q ;
\end{equation}
where $Q$ is given by
\begin{equation}
 Q = 
\left(
\begin{array}{cc}
\Phi_{00} & \Psi_{0} \\
\bar\Psi_{0} & \Phi_{00}
     \end{array}
\right) ;
\end{equation}
with
\begin{equation}
 \Phi_{00} = -\frac{1}{2} R_{ab} \ell^a \ell^b ,
\end{equation}
and
\begin{equation}
 \Psi_0 = C_{abcd} \ell^a m^b \ell^c m^d .
\end{equation}

Defining $\mathcal{X}$ by
\begin{equation}
 \mathcal{X} = 
\left(
\begin{array}{c}
\varsigma \\
\bar\varsigma
     \end{array}
\right) ;
\end{equation}
the equation for $\varsigma$ can be written as
\begin{equation}
 \ell(\mathcal{X}) = - P \mathcal{X};
\end{equation}
so that
\begin{equation}
 \ell(\ell(\mathcal{X})) = 
 - Q \mathcal{X}
;
\end{equation}
which it only involves curvature quantities.

\subsection{Approximation method for solving the geodesic deviation equation}
Although the last equation can be integrated numerically without
problems; it is sometimes convenient to have at hand some
method for approximated solutions.
So, next we present an approximation scheme that it can be applied to any
order one wishes to obtain;
although we will concentrate on the linear approximation since in weak field lens studies
it is consistent to consider linear effects of the curvature on geodesic deviations.

Let us first transform to a first order differential equation.
Defining $\mathcal{V}$ to be
\begin{equation}
 \mathcal{V} \equiv \frac{d\mathcal{X}}{d\lambda} ;
\end{equation}
and
\begin{equation}
 \mathbf{X} \equiv 
\left(
\begin{array}{c}
\mathcal{X} \\
\mathcal{V}
     \end{array}
\right) ;
\end{equation}
one obtains
\begin{equation}\label{eq:dotx}
 \ell(\mathbf{X}) = \frac{d\mathbf{X}}{d\lambda}
=
\left(
\begin{array}{c}
\mathcal{V} \\
-Q \,\mathcal{X}
     \end{array}
\right) 
= A \, \mathbf{X}
;
\end{equation}
with
\begin{equation}
 A \equiv
\left(
\begin{array}{cc}
0 & \mathbb{I} \\
-Q & 0
     \end{array}
\right) .
\end{equation}
Equation (\ref{eq:dotx}) can be reexpressed in integral form,
which gives
\begin{equation}
 \mathbf{X}(\lambda) = \mathbf{X}_0 + \int_{\lambda_0}^\lambda A(\lambda') \, \mathbf{X}(\lambda') \, d\lambda' .
\end{equation}
One can define the sequence
\begin{equation}
 \mathbf{X}_1(\lambda) = \mathbf{X}_0 + \int_{\lambda_0}^\lambda A(\lambda') \, \mathbf{X}_0 \, d\lambda'
,
\end{equation}
\begin{equation}
 \mathbf{X}_2(\lambda) = \mathbf{X}_0 + \int_{\lambda_0}^\lambda A(\lambda') \, \mathbf{X}_1(\lambda') \, d\lambda'
;
\end{equation}
and so on.

Assuming that $Q$ is in some sense small, one expects that this sequence will
converge and therefore provide for the solution.

Let us observe that
\begin{equation}
\begin{split}
 \mathbf{X}_2(\lambda) =& \mathbf{X}_1(\lambda)\\
&+ \int_{\lambda_0}^\lambda A(\lambda') \, 
\int_{\lambda_0}^{\lambda'} A(\lambda'') \, d\lambda''
 \, d\lambda' \, \mathbf{X}_0
;
\end{split}
\end{equation}
and that
\begin{equation}
 A(\lambda') \,  A(\lambda'')
= A' \,  A''
=
\left(
\begin{array}{cc}
-Q'' & 0 \\
 0 & -Q'
     \end{array}
\right)
;
\end{equation}
where we are using the notation $Q'= Q(\lambda')$.
Similarly one has
\begin{equation}
 A' \,  A''\,  A'''
=
\left(
\begin{array}{cc}
0 & -Q''  \\
Q' Q''' & 0
     \end{array}
\right)
,
\end{equation}
and
\begin{equation}
 A' \,  A''\,  A''' \,  A''''
=
\left(
\begin{array}{cc}
Q'' Q'''' & 0  \\
0 & Q' Q'''
     \end{array}
\right)
.
\end{equation}
So one can see that only at the fourth product of matrices $A$'s
one has complete second order of matrices $Q$'s.

Returning to the sequence, the third element in first order is given by
\begin{equation}
\begin{split}
 &\mathbf{X}_3(\lambda) =
\mathbf{X}_0 + \int_{\lambda_0}^\lambda 
\left(
\begin{array}{cc}
0 & \mathbb{I}  \\
-Q' & 0
     \end{array}
\right) d\lambda' \mathbf{X}_0  \\
&+
\int_{\lambda_0}^\lambda 
\int_{\lambda_0}^{\lambda'} 
\left(
\begin{array}{cc}
-Q'' & 0 \\
 0 & -Q'
     \end{array}
\right)
\, d\lambda''
\, d\lambda' \, \mathbf{X}_0 \\
&+
\int_{\lambda_0}^\lambda 
\int_{\lambda_0}^{\lambda'} 
\int_{\lambda_0}^{\lambda''} 
\left(
\begin{array}{cc}
0 & -Q''  \\
0  & 0
     \end{array}
\right)
\, d\lambda'''
\, d\lambda''
\, d\lambda' \, \mathbf{X}_0 
.
\end{split}
\end{equation}
\begin{widetext}
Working out each term, one can see that
\begin{equation}\label{eq:xlinear}
 \mathbf{X}_3(\lambda) =
\left(
\begin{array}{cc}
\mathbb{I} 
- \int_{\lambda_0}^\lambda 
\int_{\lambda_0}^{\lambda'} 
Q'' \, d\lambda'' \, d\lambda'
& 
(\lambda - \lambda_0) \mathbb{I}
- \int_{\lambda_0}^\lambda 
\int_{\lambda_0}^{\lambda'} 
(\lambda'' - \lambda_0)
Q'' \, d\lambda'' \, d\lambda'
\\
-\int_{\lambda_0}^\lambda  Q' d\lambda'
& 
\mathbb{I}
- \int_{\lambda_0}^\lambda 
(\lambda' - \lambda_0)
Q'  \, d\lambda'
     \end{array}
\right)
\mathbf{X}_0 
;
\end{equation}
\end{widetext}
where one can check that the second row is just the derivative
of the first row.

Let us note that in this equation one has not yet determined whether
the position designated by $\lambda$ is to the future or the past of the
position designated by $\lambda_0$; so that one can use this approximated
expression for both cases, keeping the same direction for the vector $\ell$.
If one changes the direction of the vector $\ell$, then one has to
take into account that $\mathcal{V}$ changes
to $-\mathcal{V}$. In particular, it is easy to see that (\ref{eq:xlinear})
is invariant under interchange of $\lambda \rightarrow -\lambda$
and $\mathcal{V} \rightarrow -\mathcal{V}$.

Note also that the double integral that appear in the first row and second 
column can by written by doing an integration by parts as
\begin{equation}
\begin{split}
\int_{\lambda_0}^\lambda 
\int_{\lambda_0}^{\lambda'} &
(\lambda'' - \lambda_0)
Q(\lambda'') \, d\lambda'' \, d\lambda'
\\
&=
\int_{\lambda_0}^{\lambda}(\lambda-\lambda')(\lambda'-\lambda_0)Q(\lambda')d\lambda'.
\end{split}
\end{equation}
In the following we will make use of this equality.

\subsection{The integrated shear and expansion}
Now in order to integrate the geodesic deviation equation, we must choose the correct initial conditions.
In the case of light rays belonging to the past null cone of the observer and intersecting 
$S$ at the source,
this initial conditions are
$\mathcal{X} = 0$  and $\mathcal{V} \neq 0$;
since one can think the beam, starts backwards in time from the observer
position, and so initially has vanishing departure, but
with nonzero expansion and shear.

Therefore in the linear approximation one has
\begin{equation}\label{eq:zangular}
\begin{split}
 \mathcal{X}(\lambda) =& 
\left( (\lambda - \lambda_0) \mathbb{I} \right. \\
&- \left. \int_{\lambda_0}^{\lambda}(\lambda-\lambda')(\lambda'-\lambda_0)Q'd\lambda'
\right)
\mathcal{V}(\lambda_0) 
;
\end{split}
\end{equation}
and
\begin{equation}
 \mathcal{V}(\lambda) = 
\left( \mathbb{I} 
- \int_{\lambda_0}^\lambda 
(\lambda' - \lambda_0)
Q'  \, d\lambda'
\right)
\mathcal{V}(\lambda_0) 
.
\end{equation}
In these integrations $\lambda_0$ indicates the position at the observer
and from now on, $\lambda_s$ will indicate the position at the source.

We observe from the first expression, that if the metric were flat ($Q=0$), in order to get a 
deviation vector constructed from $\mathcal{X}_{1}$, defined as
$\mathcal{X}$ evaluated at $\lambda_s = \lambda_0+d_s$, one must choose as initial
condition
\begin{equation}
 \mathcal{V}(\lambda_0) = \frac{1}{(\lambda_s - \lambda_0)}
\mathcal{X}(\lambda_s =\lambda_0 {+} d_s)
=
\frac{1}{d_s}
\mathcal{X}_{1}
.
\end{equation}
Let us remark that we have just fixed the scale of the affine parameter $\lambda$
to coincide with the measure of spacelike distances.

\begin{figure}[h]
\includegraphics[clip,width=0.5\textwidth]{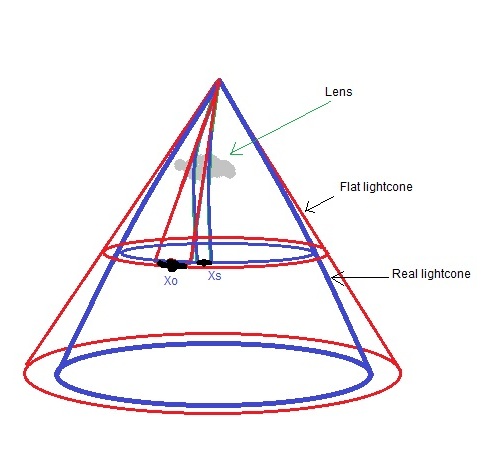}
\caption{\label{f:fig2} An object of typical dimension $d_s \varsigma^a_s$ it appears to the observer 
to have a ``size" $d_s\varsigma^a_o$.}
\end{figure}

But in the case of the presence of a gravitational lens, if an 
observer sees an image of ``size" $\mathcal{X}_o$, 
which means $\mathcal{X}_o \equiv d_s \mathcal{V}_o$ (since actually what is observed is
$\mathcal{V}_o = \mathcal{V}(\lambda_0)$~),
then it should be produced by a source of 
size $\mathcal{X}_s=\mathcal{X}(\lambda_s)$, as described by equation (\ref{eq:zangular})
and depicted in figure (\ref{f:fig2}).
 
In order to simplify the notation,
we set from now on $\lambda_0=0$ and $\lambda_s=d_s$, 
then eq.(\ref{eq:zangular}) reduces to
\begin{equation}\label{eq:zangular2}
 \mathcal{X}_s = 
\left(  \mathbb{I} 
-\frac{1}{d_s} \int_{0}^{d_s} 
\lambda'(d_s-\lambda')
Q' \, d\lambda' \right)
\mathcal{X}_o
.
\end{equation}

Note also that 
although a generic complex displacement should be $\varsigma = |\varsigma| e^{i\phi}$,
for our purposes it is enough to consider
a complex displacement $\varsigma$ of unit modulus; 
namely $\varsigma = e^{i\varphi}$. Then from equation (\ref{eq:zangular2})
one would have
\begin{widetext}
\begin{equation}
 \varsigma_s(\varphi) =
\left[ 1 - \frac{1}{d_s}\int_{0}^{d_s} 
\lambda'(d_s-\lambda')
\Phi_{00}(\lambda') \, d\lambda' -\left(   \frac{1}{d_s}\int_{0}^{d_s} 
\lambda'(d_s-\lambda')
\Psi_{0}(\lambda') \, d\lambda'\right) e^{-2 i\varphi}
\right] e^{i\varphi}
;
\end{equation}
\end{widetext}
where one can see that for the flat case one has
$\varsigma(\lambda,\varphi) =e^{i\varphi}$. From this equation, it is also observed that the expansion is only
governed by the integration of $\Phi_{00}$, and that the shear is only given by the
integration of $\Psi_{0}$.

\subsection{Expressions for the lens optical scalar in terms of Weyl and Ricci curvature from geodesic deviation equation}
In order to compare with the standard representation of the lens scalar we note that
the original deviation vector in the source will be given by eq.(\ref{eq:zangular2}), 
i.e.
\begin{equation}\label{eq:zangular3}
 \left(
\begin{array}{c}
\varsigma_{s} \\
\bar\varsigma_{s}
     \end{array}
\right)=
\left(  \mathbb{I} 
-\int_{0}^{d_s} 
\frac{\lambda'(d_s-\lambda')}{d_s}
Q' d\lambda'
\right)
\left(
\begin{array}{c}
\varsigma_{o} \\
\bar\varsigma_{o}
     \end{array}
\right)     ;
\end{equation}
if we make the following decomposition into real and imaginary part, 
\begin{eqnarray}
\varsigma_{o}&=&\varsigma_{oR}+i\varsigma_{oI} ,\\
\varsigma_{s}&=&\varsigma_{sR}+i\varsigma_{sI} ,\\
\Psi_{0}&=&\Psi_{0R}+i\Psi_{0I} ;
\end{eqnarray}
we obtain from eq.(\ref{eq:zangular3}) that
\begin{widetext}
\begin{equation}\label{eq:zetaszetao}
\begin{split}
\varsigma_{sR}=&\left(1-\int_{0}^{d_s} 
\frac{\lambda'(d_s-\lambda')}{d_s}
\left(\Phi_{00}'+\Psi_{0R}'\right)d\lambda'\right)\varsigma_{oR}-\left(\int_{0}^{d_s} 
\frac{\lambda'(d_s-\lambda')}{d_s}
\Psi_{0I}'  d\lambda'
\right) \varsigma_{oI},\\
\varsigma_{sI}=&\left(1-\int_{0}^{d_s} 
\frac{\lambda'(d_s-\lambda')}{d_s}
\left(\Phi_{00}'-\Psi_{0R}'\right) d\lambda'\right)\varsigma_{oI}
-\int_{0}^{d_s} 
\frac{\lambda'(d_s-\lambda')}{d_s}
\Psi_{0I}'d\lambda'
\varsigma_{oR}.
\end{split}
\end{equation}
\end{widetext}
Note also that in principle the integration must be made through the actual geodesic 
followed by a photon in its path from the source to observer. However the last 
expressions are valid only in the limit where the linear approximation is valid.
If one considers a linear perturbation from flat spacetime,  then
the curvature components $\Phi_{00}$ and $\Psi_0$ would be already of linear order.
Then, in the context of weak field gravitational lensing,
it is consistent to consider a null geodesic in flat spacetime;
since the actual null geodesic can be thought as 
a null geodesic in flat spacetime plus some corrections of higher orders.

We choose then, a null geodesic coming from a source located at a distance $d_s$ from the 
observer, and select a Cartesian coordinate system where
this geodesic will propagate along the $y$  negative direction.
As we mentioned previously, one can actually integrate the equations either along the physical 
direction or one can integrate to the past from a null geodesic that starts at the observer
position. We make this second choice.

We also need a null tetrad $\{l^a,m^a,\bar{m}^a,n^a\}$, adapted to this geodesic,
\begin{equation}\label{eq:tetradflat}
\begin{split}
l^a=&(-1,0,1,0),\\
m^a=&\frac{1}{\sqrt{2}}(0,i,0,1),\\
\bar{m}^a=& \frac{1}{\sqrt{2}}(0,-i,0,1),\\
n^a=&\frac{1}{2}(-1,0,-1,0).
\end{split}
\end{equation}

Now, in order to compare with the usual expressions for the lens scalars 
$\kappa, \gamma_1$ and $\gamma_2$, 
let us recall that they are defined via the relation eq.(\ref{eq:standarlensequation});
but since it is a linear relation, one can relate the deviation vectors by the same
matrix, namely
\begin{equation}\label{eq:deltabeta}
\varsigma^i_s=A^i_j\varsigma^j_o;
\end{equation}
where $\{\varsigma^i_s,\,\varsigma^i_o\}$ are the spatial vector associated with
$\{\varsigma_s,\,\varsigma_o\}$ respectively.
In this expression,
it is needed to determine the meaning of the indices $(i,j)$ of the two dimensional space
of the images. In order to observe the natural Cartesian orientation we identify the first
component of the two dimensional space with the $z$ one of the complete system, 
and the second component of the two dimensional space with the $x$ one.
We need then, to know the components of the spatial vectors $\varsigma^a_o$ generated 
by $\varsigma_o$ and similarly by $\varsigma_s$ in a Cartesian like coordinate system.
In the case of $\varsigma^a_o$, it is given by
\begin{equation}
\begin{split}
\varsigma^a_o=&\varsigma_o \bar{m}^a+\bar\varsigma_o{m}^a\\
=&\frac{1}{\sqrt{2}}\left(\varsigma_o (0,-i,0,1)+\bar\varsigma_o (0,i,0,1)\right)\\
=&\frac{1}{\sqrt{2}}\left(0,i(\bar{\varsigma}_o-\varsigma_o),0,(\varsigma_o+\bar{\varsigma}_o)\right)\\
=&\frac{2}{\sqrt{2}}\left(0,\varsigma_{oI},0,\varsigma_{oR}\right);
\end{split}
\end{equation} 
and a similar expression is obtained for $\varsigma^a_s$.

Therefore, by replacing into eq.(\ref{eq:deltabeta}), we obtain
\begin{eqnarray}
\varsigma_{sR}&=&(1-\kappa-\gamma_1 )\varsigma_{oR}-\gamma_2\,\varsigma_{oI},\\
\varsigma_{sI}&=&-\gamma_2\,\varsigma_{oR}+(1-\kappa+\gamma_1)\varsigma_{oI};
\end{eqnarray}
which by comparing with eq.(\ref{eq:zetaszetao}), implies that
\begin{eqnarray}\label{eq:lensscalars}
\kappa&=&\frac{1}{d_s}\int_{0}^{d_s} 
\lambda'(d_s-\lambda')
\Phi_{00}'\, d\lambda',\\
\gamma_1&=&\frac{1}{d_s}\int_{0}^{d_s} 
\lambda'(d_s-\lambda')
\Psi_{0R}'\, d\lambda',\\
\gamma_2&=&\frac{1}{d_s}\int_{0}^{d_s} 
\lambda'(d_s-\lambda')\Psi_{0I}'\, d\lambda'.
\end{eqnarray}
Let us emphasize that these expressions for the weak field lens quantities are explicitly
gauge invariant, since they are given in terms of the curvature components,
which are gauge invariant. This is in contrast to the usual treatment of weak field gravitational lensing
found in the literature, which use for example equation (2.17) of reference
\cite{Schneider92} as the source for the calculation of the lens scalars.

Note that the last two equations can be written as
\begin{equation}
\gamma_1+i\gamma_2 = \frac{1}{d_s}\int_{0}^{d_s} 
\lambda'(d_s-\lambda')
\Psi_{0}'\, d\lambda'.
\end{equation}
As a final comment to this section it is important to remark that these expressions are 
valid for any weak field gravitational lens on a perturbed flat spacetime, without 
restriction on the size of the lens compared with the other distances.
However, if we make use of the hypothesis of thin lens, these equations 
can be further simplified, as we will show below. 

\section{The axially symmetric lens (including the spherically symmetric case)}\label{sec:axisymm}
When one observes an astrophysical system very often one needs to extract information
of the bulk of the matter distribution, which normally involves to make some 
model assumptions on the nature of the distribution. So, very often
one considers spherically symmetric models or the less restrictive case
of axially symmetric distribution. In this later case the axis coincide with
the line passing through the central region of the distribution and
the observer.

In this section then we consider the case of an axially symmetric gravitational lens
without introducing further assumptions on the extend of the lens. Later we will
consider thin lenses.

Using the same setting as in the last section, one is studying the motion of a photon
which travels along the negative $y$ direction, with impact parameter $J$ and angle
$\vartheta$ from the $z$ axis. Then one notes that the component $\Phi_{00}$ is 
a spin zero real quantity,
and it only depends on the $(J,y)$ coordinates. While the component $\Psi_0$ is 
a spin two complex quantity and it
has the functional dependence
\begin{equation}
 \Psi_0 = |\Psi_0| e^{2 i \vartheta + \text{phase}} ;
\end{equation}
where the phase is gauge dependent. For reasons that will become more clear during
the study of spherically symmetric systems, we define the real quantities $\psi_0(J,y)$
from
\begin{equation}
 \Psi_0(J,y,\vartheta) = - \psi_0(J,y) e^{2 i \vartheta} .
\end{equation}

From this we deduce that the optical scalars have the following dependence
\begin{equation}
\kappa(J) = \frac{1}{d_s}\int_{0}^{d_s} 
\lambda'(d_s-\lambda')
\Phi_{00}(J,\lambda')\, d\lambda', 
\end{equation}
and
\begin{equation}
\gamma_1+i\gamma_2 = - \frac{1}{d_s}  e^{2 i \vartheta}  \int_{0}^{d_s} 
\lambda'(d_s-\lambda')
\psi_{0}(J,\lambda')\, d\lambda' .
\end{equation}
This invite us to also define the real quantity $\gamma(J,y)$ from
\begin{equation}\label{eq:gama1masgama2}
 \gamma_1+i\gamma_2 = - \gamma e^{2 i \vartheta} ;
\end{equation}
so that one simply has
\begin{equation}
 \gamma(J,y)  = \frac{1}{d_s}\int_{0}^{d_s} \lambda'(d_s-\lambda')
\psi_{0}(J,\lambda')\, d\lambda' .
\end{equation}

\section{The thin lens approximation}\label{sec:thin}
\subsection{The general case}
Now, we will consider the case of a lens whose size is small compared with the distances to 
the source and the observer.
Let there be again a Cartesian coordinate system 
such that the lens can be thought to be localized around the plane $y=0$. 

Then as it was indicated in the last section, in the linear approximation we can replace the 
actual null geodesic by one in a flat spacetime. 
Then, considering a null geodesic as in the previous section, coming from a source located 
at a distance $d_s$ from the observer, and at a distance $d_{ls}$ from the lens, 
coming parallel to the $y$ axis, but in the negative direction;
we will use $J$ to represent the impact parameter and  $\vartheta$
to denote the angle of the trajectory as measured from the $z$ axis, in the $(z,x)$ plane.
We choose the scale of the affine parameter $\lambda$ such that the geodesic is described by 
\begin{equation}
(x(\lambda), y(\lambda),z(\lambda))=(x_0,\lambda-d_{l},z_0);
\end{equation}
i.e., $\lambda=0$ indicates the position of the observer, and $\lambda=d_s$ the 
position of the source.

Then if we represent generically by $C$  each one of the scalars $\{\Phi_{00},\Psi_{0}\}$, that 
appear in the expressions for the lens scalars, we have that by doing an integration by parts
one obtains the relation
\begin{equation}
\begin{split}
\int_0^{d_s}\lambda'(d_s-\lambda')C(\lambda')d\lambda' =& \\
\left.\lambda'(d_s-\lambda')\tilde C(\lambda')\right|^{d_s}_0
&-\int^{d_s}_0(d_s-2\lambda')\tilde C(\lambda')d\lambda' \\
=&
-\int^{d_s}_0(d_s-2\lambda')\tilde C(\lambda')d\lambda' ;
\end{split}
\end{equation}
where 
\begin{equation}
\tilde C(\lambda')=\int^{\lambda'}_0 C(\lambda'')d\lambda''.
\end{equation}

Then if we assume a thin lens, the scalars $C$, will be sharply peaked around $\lambda=d_l$, 
where it is located, and $\tilde C(\lambda)$ can be approximated by 
\begin{eqnarray}\label{eq:c-hat}
\tilde C(\lambda') \cong \left\{\begin{array}{ll}
0 & \quad \forall \, \lambda < d_l - \delta \\
\hat C & \quad \forall \,\lambda\geq d_l + \delta ;
\end{array}\right.
\end{eqnarray}
where $\delta \ll d_l$, $\delta \ll d_{ls}$ and $\delta \ll d_s$.
Therefore, we obtain
\begin{equation}
\begin{split} 
\int^{d_s}_0(d_s-2\lambda')\tilde C(\lambda')d\lambda'&\;\cong
\; \hat C\int^{d_s}_{d_l}(d_s-2\lambda')d\lambda'\\
&= \, \hat C d_l(d_l-d_s)\\
&=-\hat C d_ld_{ls} ;
\end{split}
\end{equation}
where we have neglected terms of order $O(\frac{\delta}{d_l})$.
Obviously all this also holds for the particular case of a delta Dirac distribution for
the curvature components; however our relaxed notion of thin lens is entirely expressed by
the (\ref{eq:c-hat}) behavior.

Finally we conclude that in the thin lens approximation, the expressions for the lens 
scalars are reduced to
\begin{equation}\label{eq:g1masig}
\begin{split}
\kappa = \frac{d_ld_{ls}}{d_s}\hat{\Phi}_{00},
\end{split}
\end{equation}
\begin{equation}
\begin{split}\label{eq:g1masig2}
\gamma_1+i\gamma_2 = \frac{d_ld_{ls}}{d_s}\hat{{\Psi}}_0,
\end{split}
\end{equation}
where
\begin{equation}
\begin{split}
\hat{\Phi}_{00}=&\int^{d_s}_0\Phi_{00}d\lambda,\\
\hat{\Psi}_{0}=&\int^{d_s}_0\Psi_{0}d\lambda,
\end{split}
\end{equation}
are the projected curvature scalars along the line of sight.

We again emphasize that these expressions for the lens scalars are explicitly 
gauge invariants.

\subsection{The axially symmetric case (which includes the spherically symmetric case)}
\subsubsection{The lens scalars  in terms of projected Ricci and Weyl Scalars}
For axially symmetric lens (and in fact spherically symmetric lens), 
the projected curvature scalars are given by
\begin{eqnarray}
\hat\Phi_{00}(J)&=&\int^{d_s}_0\Phi_{00}(\lambda')d\lambda',\\
\hat{\Psi}_{0}(J)&=&-e^{2i\vartheta}\hat{\psi}_{0}(J).
\end{eqnarray}
where one can see that
\begin{eqnarray}
 \hat{\psi}_{0}(J) &=&-e^{-2i\vartheta}\int^{d_s}_{0} \Psi_0(\lambda')d\lambda'.
 \end{eqnarray}
The reason for the minus sign choice is that in many common astrophysical situations
one would find $\hat{\psi}_{0}(J) > 0$.

By replacing in eqs. (\ref{eq:g1masig}) and (\ref{eq:g1masig2}), we obtain for the lens scalars
\begin{eqnarray}\label{eq:lensscalarthin}
\kappa &=&\frac{d_{ls} d_l}{d_s}\hat{\Phi}_{00}(J),\\
\gamma_1 &=& - \frac{d_{ls} d_l}{d_s}\hat{\psi}_{0}(J)\cos(2\vartheta),\label{eq:lensscalarthin1}\\
\gamma_2 &=& - \frac{d_{ls} d_l}{d_s}\hat{\psi}_{0}(J)\sin(2\vartheta) ,\label{eq:lensscalarthin2}
\end{eqnarray}
which implies that
\begin{equation}\label{eq:lensscalarthin-g}
\gamma =  \frac{d_{ls} d_l}{d_s}\hat{\psi}_{0}(J) .
\end{equation}
This equations can be compared to those of reference \cite{Frittelli00};
where they use different notation but similar content.

\subsubsection{Deflection angle in terms of projected Ricci and Weyl Scalars}
We wish now to express the deflection angle in terms of the curvature scalars. 

From eq.(\ref{eq:standarlensequation}) we know that
\begin{equation}
A^i_j=\frac{d\beta^i}{d\theta^j}=\delta^i_j-\frac{d_{ls}}{d_s}\frac{d\alpha^i}{d\theta^j}
=\delta^i_j-\frac{d_{ls}d_l}{d_s}\frac{d\alpha^i}{dx^j};
\end{equation}
where in the last equality we have used that in the thin lens approximation 
$\frac{d}{d\theta^i}\approx d_l\frac{d}{dx^i}$.

We define the components of $\alpha^i=(\alpha^1,\alpha^2)$ as
\begin{equation}
(\alpha^i)= \alpha(J) (\frac{z_0}{J}, \frac{x_0}{J} );
\end{equation}
since as we mentioned above we are respecting the Cartesian orientation in the two
dimensional space of the images.
We then obtain that the shears and convergence can be written as,
\begin{eqnarray}
\kappa&=&\frac{1}{2}\frac{d_{ls}d_l}{d_s}
\left(\frac{d\alpha^1}{dz_0}+\frac{d\alpha^2}{dx_0}\right),\\
\gamma_1&=& \frac{1}{2}\frac{d_{ls}d_l}{d_s}
\left(\frac{d\alpha^1}{dz_0}-\frac{d\alpha^2}{dx_0}\right),\\
\gamma_2&=&\frac{d_{ls}d_l}{d_s}\frac{d\alpha^1}{dx_0}=\frac{d_{ls}d_l}{d_s}\frac{d\alpha^2}{dz_0}.
\end{eqnarray}

Noting that 
\begin{equation}
\begin{split}
x_0=&J\sin(\vartheta),\\
 z_0=&J\cos(\vartheta);
\end{split}
\end{equation} 
we obtain
\begin{eqnarray}\label{eq:scalarlensalpha}
\kappa &=&\frac{1}{2}\frac{d_{ls}d_l}{d_s}
\left( \frac{d\alpha}{dJ} +  \frac{\alpha(J)}{J}  \right),\\
\gamma_1 &=& \frac{1}{2}\frac{d_{ls}d_l}{d_s} \label{eq:scalarlensalpha1}
\,\cos(2\vartheta) \, \left( \frac{d\alpha}{dJ} -  \frac{\alpha(J)}{J}  \right),\\
\gamma_2 &=&\frac{1}{2}\frac{d_{ls}d_l}{d_s} \label{eq:scalarlensalpha2}
\,\sin(2\vartheta) \, \left( \frac{d\alpha}{dJ} -  \frac{\alpha(J)}{J}  \right).
\end{eqnarray}

It is interesting to note that
\begin{equation}
\kappa-
\gamma_1 \cos(2\vartheta) - \gamma_2 \sin(2\vartheta) 
=\frac{d_ld_{ls}}{d_{s}}\frac{\alpha(J)}{J};
\end{equation}
from which, using  eqs. (\ref{eq:lensscalarthin}-\ref{eq:lensscalarthin2}), it is deduced that
\begin{equation}\label{eq:simplealpha}
\alpha(J)=
J(\hat{\Phi}_{00}(J)+\hat{\psi}_{0}(J)).
\end{equation}
It is worthwhile to remark that this constitutes an equation for the bending angle
expressed in terms of the gauge invariant curvature components in a very simple
compact form. We do not have knowledge of a previous presentation of this equation.

It is also important to emphasize that we have derived the expression for the deflection
angle  from the information contained in the calculation of the optical scalars,
coming from the geodesic deviation equation.

Note that, at first sight, it seems that if we reconstruct the lens scalars using eqs.
(\ref{eq:scalarlensalpha}-\ref{eq:scalarlensalpha2}), 
from expression   (\ref{eq:simplealpha})   for the deflection angle, one would
obtain some condition on the bending angle when compared with eqs. 
(\ref{eq:lensscalarthin}-\ref{eq:lensscalarthin2}); 
however this is only an apparent inconsistence.
Let us see this in more detail. To begin with, by replacing eq.
(\ref{eq:simplealpha}) into eqs.
(\ref{eq:scalarlensalpha}-\ref{eq:scalarlensalpha2}), one obtains
\begin{equation}
\begin{split}
\label{eq:scalarssinbianchi}
\kappa=\frac{d_{ls}d_l}{2 d_s} &\left[ 2\left(\hat{\Phi}_{00}(J)+\hat{\psi}_{0}(J)\right) \right.\\
 & \left.\quad + J\frac{d (\hat{\Phi}_{00}+\hat{\psi}_{0})}{d J}\right],
\end{split}
\end{equation}
\begin{equation}\label{eq:scalarssinbianchi1}
\gamma_1 = \frac{d_{ls}d_l}{2 d_s}
\,\cos(2\vartheta) \, J
\left(\frac{d \hat{\Phi}_{00}}{d J} 
+\frac{d \hat{\psi}_{0}}{d J}\right), 
\end{equation}
\begin{equation}\label{eq:scalarssinbianchi2}
\gamma_2 = \frac{d_{ls}d_l}{2 d_s} \sin(2\vartheta)
J\left(\frac{d \hat{\Phi}_{00}}{d J} 
+\frac{d \hat{\psi}_{0}}{d J}\right)
.
\end{equation}

Proceeding with the calculation, we now use one of the Bianchi identities, as expressed in the
GHP formalism\cite{Geroch73}, namely
\begin{equation}
\text{\Thorn}\Psi_1-\eth'\Psi_0+\eth\Phi_{00} - \text{\Thorn}\Phi_{01}= 0;
\end{equation}
where in our case $\text{\Thorn}=l^a\partial_a$, $\eth=m^a\partial_a$, and $\eth'=\bar{m}^a\partial_a$, i.e.
\begin{equation}
\begin{split}
\text{\Thorn}=&-\frac{\partial}{\partial t}+\frac{\partial}{\partial y},\\
\eth=&\frac{1}{\sqrt{2}}\left(i\frac{\partial}{\partial x}+\frac{\partial}{\partial z}\right),\\
\eth'=&\frac{1}{\sqrt{2}}\left(-i\frac{\partial}{\partial x}+\frac{\partial}{\partial z}\right);
\end{split}
\end{equation}
that is, we are using the flat null tetrad system.
The expression for the edth operator is correct due to the fact that the intrinsic
two dimensional metric in the space $(x, z)$ is constant and therefore $\eth= m$.

Let us now change to a polar coordinate system in the two dimensional subspace,
so that 
\begin{equation}
\begin{split}
\frac{\partial}{\partial x}=&\frac{\partial J}{\partial x}\frac{\partial}{\partial J}+\frac{\partial \vartheta}{\partial x}\frac{\partial}{\partial \vartheta},\\
\frac{\partial}{\partial z}=&\frac{\partial J}{\partial z}\frac{\partial}{\partial J}+\frac{\partial \vartheta}{\partial z}\frac{\partial}{\partial \vartheta};
\end{split}
\end{equation}
with 
\begin{equation}
\begin{split}
\frac{\partial \vartheta}{\partial x}=&\frac{\cos(\vartheta)}{J},\\
\frac{\partial \vartheta}{\partial z}=&-\frac{\sin(\vartheta)}{J} .
\end{split}
\end{equation}

In this case the metric of the two dimensional space $(J,\vartheta)$ is not constant, so
that in principle the edth operator acting on a quantity $f$ of type $(p,q)$ should
be\cite{Geroch73}
\begin{equation}
\eth f = m(f) + (-p\beta+q\beta')f; 
\end{equation}
but a direct calculation in the $(J,\vartheta)$ frame gives all spin coefficients
zero. Therefore in this frame we also have
$\eth f = m(f).$
Then we get
\begin{equation}
\begin{split}
\eth=&
\frac{1}{\sqrt{2}}e^{i\vartheta}\left(\frac{\partial}{\partial J}+\frac{i}{J}\frac{\partial}{\partial \vartheta}\right),\\
\eth'=&\frac{1}{\sqrt{2}}e^{-i\vartheta}\left(\frac{\partial}{\partial J}-\frac{i}{J}\frac{\partial}{\partial \vartheta}\right).
\end{split}
\end{equation}

If we now project the Bianchi identity on the line of sight direction, i.e. 
by integrating along the $y$-direction, we obtain
\begin{equation}
\Psi_1 |^{d_s}_{0} - \eth'\hat{\Psi}_{0} + \eth\hat{\Phi}_{00} -\Phi_{01} |^{d_s}_{0} = 0;
\end{equation}
which, assuming $\Psi_1\approx 0$ and $\Phi_{01}\approx 0$ far away from the lens, it implies
\begin{equation}\label{eq:proyBianchi}
\eth'(\hat{\psi}_{0}e^{2i\vartheta})=-\eth(\hat{\Phi}_{00}).
\end{equation} 
From this one finds
\begin{equation}
\frac{d \hat{\psi}_{0}}{d J} + 2\frac{\hat{\psi}_{0}}{J} =
-\frac{d \hat{\Phi}_{00}}{d J} .
\end{equation}
Then by replacing this relation into eqs. (\ref{eq:scalarssinbianchi}-\ref{eq:scalarssinbianchi2}) 
we obtain eqs. (\ref{eq:lensscalarthin}-\ref{eq:lensscalarthin2}), as anticipated.

The Bianchi identities have not been used very often in the context of gravitational lenses;
however we note that in reference \cite{Kling05,Kling08} they
have used them to obtain a Poisson like equation in order to determine the 
matter distribution.

For the study of
the errors committed in the use of the thin
lens approximation one can read \cite{Kling08b}.

\section{Detailed study of stationary spherically symmetric lenses}\label{sec:spher}

Up to now we have presented gauge invariant expressions for the deviation angle and
the optical scalars in terms of the curvature components of the null tetrad adapted to
the motion of the photons.
In order to obtain expressions that use information of the structure of the sources
one has to work with frames adapted to the geometry of the matter distribution which
forms the gravitational lens.
Therefore in this section we study the case of stationary spherically symmetric sources.

\subsection{Spacetime geometry in standard coordinate system}

\subsubsection{The metric}

For stationary spherically symmetric spacetime, the line element can be expressed by
\begin{equation}\label{eq:ds1}
 ds^2 = a(r) dt^2 - b(r) dr^2 - r^2(d\theta^2 + \sin^2 \theta d\varphi^2) .
\end{equation}

It is convenient to define $\Phi(r)$ and $M(r)$ from
\begin{equation}
 a(r) = e^{2 \Phi(r)} ,
\end{equation}
and
\begin{equation}
b(r) = \frac{1}{1 - \frac{2 M(r)}{r} }
.
\end{equation}

The more general distribution of energy-momentum compatible with spherical 
symmetry is described by an energy-momentum tensor given by
\begin{equation}
 T_{tt} =  \varrho e^{2 \Phi(r)};
\end{equation}
\begin{equation}
 T_{rr} =    \frac{P_r}{\left( 1 - \frac{2 M(r)}{r} \right) } ;
\end{equation}
\begin{equation}
 T_{\theta \theta} =    P_t \, r^2 ;
\end{equation}
\begin{equation}
 T_{\varphi \varphi} =    P_t \,  r^2 \sin(\theta)^2 ;
\end{equation}
where we have introduced the notion of radial component  $P_r$ and
tangential component  $P_t$.

The Einstein field equations
\begin{equation}
G_{ab} = -8\pi T_{ab} ,
\end{equation} 
in terms of the previous variables are
\begin{equation}\label{eq:rho}
 \frac{dM}{dr} = 4\pi r^2 \varrho ,
\end{equation}
\begin{equation}\label{eq:mgrt}
r^2\frac{d\Phi}{dr} = \frac{M + 4\pi r^3 P_r}{1 - \frac{2 M(r)}{r} } ,
\end{equation}
\begin{equation}
\begin{split}\label{eq:pt}
 r^3 &\left(\frac{d^2\Phi}{dr^2}+(\frac{d\Phi}{dr})^2\right)( 1-\frac{2 M}{r})\\
&+ r^2 \frac{d\Phi}{dr} (1 - \frac{M}{r}-\frac{dM}{dr})\\
&- r \frac{dM}{dr} + M 
= 8\pi r^3 P_t  .
\end{split}
\end{equation}
The conservation equation is 
\begin{equation}\label{eq:consrt}
 \frac{dP_r}{dr} = -(\varrho + P_r ) \frac{d\Phi}{dr} 
- \frac{2}{r} (P_r - P_t)  .
\end{equation}

\subsection{Geometry with respect to a null system}

\subsubsection{The tetrad}
For our purpose, it is more convenient to use a null coordinate system to describe the spherically symmetric geometry.
Let us introduce then, a function
\begin{equation}
 u = t - r^* ;
\end{equation}
where $r^*$ is chosen so that $u$ is null. Then by inspection
of equation (\ref{eq:ds1}) one can see that
\begin{equation}\label{eq:du}
 du = dt - \frac{dr^*}{dr} dr = dt - \sqrt{\frac{b}{a}} dr ;
\end{equation}
since then one has
\begin{equation}\label{eq:ds2}
 ds^2 = a\, du^2 + 2 \sqrt{a b} du dr 
        - r^2(d\theta^2 + \sin^2 \theta d\varphi^2) .
\end{equation}

It is natural to define the principal null direction $\tilde\ell_P$
from
\begin{equation}\label{eq:ell}
 \tilde\ell_P = du ;
\end{equation}
which implies that the vector is
\begin{equation}
 \tilde\ell_P^a = g^{ab} du_b = \frac{1}{\sqrt{a b}} 
\left( \frac{\partial}{\partial r}\right)^a ;
\end{equation}
where we have used that
\begin{equation}\label{eq:invds2}
\begin{split}
 (g^{ab}) =& 
\frac{2}{\sqrt{a b}} \frac{\partial}{\partial u}
                          \frac{\partial}{\partial r}
- \frac{1}{b} \frac{\partial}{\partial r}\frac{\partial}{\partial r}\\
-& \frac{1}{r^2}
(\frac{\partial}{\partial \theta}\frac{\partial}{\partial \theta}
+\frac{1}{\sin^2 \theta}   
\frac{\partial}{\partial \varphi}\frac{\partial}{\partial \varphi}
) .
\end{split}
\end{equation}

Let us define the null tetrad
\begin{equation}
 \tilde\ell_P = \mathbb{A} \frac{\partial}{\partial r} ,
\end{equation}
\begin{equation}
 \tilde{n}_P = \frac{\partial}{\partial u} + U \mathbb{A} \frac{\partial}{\partial r} ,
\end{equation}
with the complex null vector
\begin{equation}
 \tilde{m}_P = \frac{\sqrt{2} P_0}{r}  \frac{\partial}{\partial \zeta} ;
\end{equation}
in terms of the stereographic coordinate $\zeta$.

Therefore, one has
\begin{equation}
 \mathbb{A} = \frac{1}{\sqrt{a b}} ,
\end{equation}
and
\begin{equation}
 U = - \frac{1}{2 b \mathbb{A}^2} = -\frac{a}{2}
.
\end{equation}

It is worthwhile to note that we have chosen to keep using $r$ as a coordinate;
which measures the surfaces of the symmetric spheres.
Instead one could have chosen to use an affine coordinate $\tilde r$ so that
one would have $ \tilde{\ell} =  \frac{\partial}{\partial \tilde r}$; but then the surfaces of
the symmetric spheres would be some function of $\tilde r$, different from $4\pi \tilde r^2$.

\subsection{The spin coefficients scalars and curvature components}
For the spherically symmetric metric, the non vanishing spin coefficients are
\begin{eqnarray}
\tilde\rho&=&-\frac{\mathbb{A}}{r},\\
\tilde\rho'&=&-\frac{U\mathbb{A}}{r},\\
\beta &=& \frac{1}{\sqrt{2} r} \left( -\frac{\partial P_0}{\partial y_2} + i\frac{\partial P_0}{\partial y_3} \right),\\
\beta' &=& \frac{1}{\sqrt{2} r} \left( -\frac{\partial P_0}{\partial y_2} - i\frac{\partial P_0}{\partial y_3} \right),\\
\tilde\epsilon'&=&\frac{1}{2}\mathbb{A}\frac{d\mathbb{U}}{dr};
\end{eqnarray}
where we are using for the stereographic coordinate, the decomposition
$\zeta = \frac{1}{2}(y_2 + i y_3)$.

The curvature components that are different from zero are:
\begin{equation}
 \tilde{\Phi}_{00} = - \frac{\mathbb{A}}{r} \frac{d\mathbb{A}}{dr} ,
\end{equation}
\begin{equation}
\begin{split}
\tilde{\Phi}_{11} =&
 -\frac{1}{4} \frac{d\mathbb{A}}{dr} \frac{dU}{dr} \mathbb{A}
- \frac{1}{4} \nabla^2(U) \mathbb{A}^2 \\
&+ \frac{1}{2 r}  \frac{dU}{dr} \mathbb{A}^2
+ \frac{1}{r^2} (\frac{1}{2} \mathbb{A}^2 U + \frac{1}{4}) ,
\end{split}
\end{equation}
\begin{equation}
 \tilde{\Phi}_{22} =  - \frac{\mathbb{A }U^2}{r} \frac{d\mathbb{A}}{dr} ,
\end{equation}
\begin{equation}
\begin{split}
\tilde{\Lambda} =& \frac{1}{12} \frac{d\mathbb{A}}{dr} \frac{dU}{dr} \mathbb{A}
+ \frac{1}{12} \nabla^2(U) \mathbb{A}^2 \\
&+ \frac{1}{r} \left( \frac{1}{3} \frac{d\mathbb{A}}{dr} \mathbb{A} U
+\frac{1}{6}\frac{dU}{dr} \mathbb{A}^2 \right)\\
&+ \frac{1}{r^2} (\frac{1}{6} \mathbb{A}^2 U + \frac{1}{12} ) ,
\end{split}
\end{equation}
and
\begin{equation}
\begin{split}
 \tilde{\Psi}_2 =&
 -\frac{1}{6} \frac{d\mathbb{A}}{dr} \frac{dU}{dr} \mathbb{A}
- \frac{1}{6} \nabla^2(U) \mathbb{A}^2 \\
&+ \frac{1}{r} \left( \frac{1}{3} \frac{d\mathbb{A}}{dr} \mathbb{A} U
+\frac{2}{3}\frac{dU}{dr} \mathbb{A}^2 \right)\\
&+ \frac{1}{r^2} (-\frac{1}{3} \mathbb{A}^2 U - \frac{1}{6} ) .
\end{split}
\end{equation}

Note that from (\ref{eq:du}) and (\ref{eq:ell}) one has that
\begin{equation}
 \tilde\ell_P = dt - \sqrt{\frac{b}{a}} dr ,
\end{equation}
and therefore
\begin{equation}
 \tilde\ell^a_P = \frac{1}{a} (\frac{\partial}{\partial t})^a
+ \sqrt{\frac{1}{a b}} \left.(\frac{\partial}{\partial r})\right|_t^a .
\end{equation}

Also let us note that
\begin{equation}
 \frac{\partial}{\partial u} = \frac{\partial}{\partial t} ;
\end{equation}
which then implies that
\begin{equation}
\begin{split}
 \tilde{n}_P =&\frac{1}{2}
\frac{\partial}{\partial t}
- \frac{1}{2 }\sqrt{\frac{a}{b}} \left.(\frac{\partial}{\partial r})\right|_t^a
.
\end{split}
\end{equation}

In these last equations $\left.(\frac{\partial}{\partial r})\right|_t^a$ is meant at
constant $t$; as opposite to the previous equations in which  $\frac{\partial}{\partial r}$ was meant
at constant $u$.

\subsection{Spinor Ricci components in terms of energy-momentum components in the non-isotropic case}

The spinor Ricci components can be written in terms of the energy-momentum distribution as
\begin{eqnarray}\label{eq:phis}
 \tilde{\Phi}_{00} &=&\frac{4 \pi}{ a}\left(  
 \varrho + P_r \right),\\
\tilde{\Phi}_{11} &=& \pi\left(  \varrho - P_r + 2  P_t \right),\label{eq:phis11}\\
 \tilde{\Phi}_{22} &=& a \pi \left(  \varrho + P_r \right),\\
\tilde{\Lambda} &=& \frac{\pi}{3}(\varrho -  P_r - 2 P_t);
\end{eqnarray}
These expressions are exact for the spherically symmetric spacetime. If one needs linear 
expressions around flat spacetime, one must set $a=1$.

Note that one has
\begin{equation}
 \tilde{\Phi}_{22} = U^2 \tilde\Phi_{00} .
\end{equation}

Using the expressions for $\tilde{\Phi}_{11}$ and $\tilde{\Lambda}$ one can prove that
\begin{equation}\label{eq:fi11m3lam-b}
\begin{split}
\tilde{\Phi}_{11} + 3 \tilde{\Lambda} =&\frac{\mathbb{A}}{r} \frac{d(U \mathbb{A})}{dr}
+ \frac{1}{r^2} (U \mathbb{A}^2 + \frac{1}{2})\\
=& 2 \pi (\varrho - P_r)
.
\end{split}
\end{equation}

Also, from the relation of the null tetrad components with
the old variables, one can obtain that
\begin{equation}\label{eq:ua2}
 U \mathbb{A}^2 + \frac{1}{2} = \frac{M(r)}{r} .
\end{equation}
This equation gives $U$ in terms of $\mathbb{A}$ and $M$. 

Using this in the expression for $\tilde{\Phi}_{00}$ one obtains
\begin{equation}
 \frac{1}{\mathbb{A}} \frac{d\mathbb{A}}{dr} =
\frac{4 \pi r (\varrho + P_r)}{\left( \frac{2 M}{r} - 1\right) }
.
\end{equation}
This is a useful equation only involving $\mathbb{A}$; which allows it's calculation
in terms of the components of the energy-momentum tensor.

 The contracted Bianchi identity (2.37) of \cite{Geroch73} for spherically symmetric metrics is
\begin{equation}
\text{\Thorn}\tilde{\Phi}_{11}+\text{\Thorn}'\tilde{\Phi}_{00}+3\text{\Thorn}\tilde{\Lambda}=(\tilde{\rho}'+\tilde{\bar{\rho}}')\tilde{\Phi}_{00}+2(\tilde{\rho}+\tilde{\bar{\rho}})\tilde{\Phi}_{11}.
\end{equation}
or explicitely,
\begin{equation}
\begin{split}
\frac{d\tilde{\Phi}_{11}}{dr}+&\mathbb{A}U\frac{d\tilde{\Phi}_{00}}{dr}+2\frac{dU}{dr}\mathbb{A}\tilde{\Phi}_{00}+3\frac{d\tilde{\Lambda}}{dr} \\
=& -2\frac{\mathbb{A}U}{r}\tilde{\Phi}_{00}-4\frac{\mathbb{A}}{r}\tilde{\Phi}_{11} ,
\end{split}
\end{equation}
which gives the conservation
equation in the form
\begin{equation}\label{eq:dP}
\frac{dP_r}{dr} = -(\varrho + P_r )\frac{m_g(r)}{r^2} - \frac{2}{r}(P_r - P_t)
 ;
\end{equation}
where we are using
\begin{equation}\label{eq:mg2}
 m_g(r) = \frac{r^2}{2} \frac{d\ln U}{dr} .
\end{equation}

\subsection{Simple relation for Weyl component $\tilde{\Psi}_2$}
Let us observe that
\begin{equation}\label{eq:psi2lambda}
 \tilde{\Psi}_2 + 2 \tilde{\Lambda} =
\frac{\mathbb{A}}{r} \frac{d(\mathbb{A} U)}{dr}
.
\end{equation}
Then from equation (\ref{eq:fi11m3lam-b}) one can deduce that
\begin{equation}\label{eq:psi2rhoP}
\begin{split}
 \tilde{\Psi}_2  
=& \frac{4 \pi}{3} (\varrho - P_r  + P_t)-\frac{M}{r^3} .
\end{split}
\end{equation}
This is a very simple relation for $\tilde{\Psi}_2(r)$ in terms of the energy
density $\varrho(r)$, the spacelike components  and the mass function $M(r)$.
Our expression generalizes those of reference \cite{Dyer77} for the
case of anisotropic energy-momentum tensor.

\subsection{The bending angle and lens scalars in terms of energy-momentum components,
curvature components and $M(r)$}
\subsubsection{Relation between the scalars curvatures in the two different tetrads}
In order to express the function $\alpha(J)$ in terms of the curvature scalars defined with 
the spherically symmetric tetrad, we need to know how the tetrads transform between them. 
To do so, let us recall that at linear order, we only need the transformation between 
the flat tetrad $\{l^a,m^a,\bar{m}^a,n^a\}$ adapted to the null geodesic coming from the 
source, and a flat tetrad $\{\tilde{l}^a,\tilde{m}^a,\bar{\tilde{m}}^a,\tilde{n}^a\}$ obtained 
from $\{\tilde{l}^a_P,\tilde{m}^a_P,\tilde{\bar{m}}^a_P,\tilde{n}^a_P\}$ by setting 
$a=b=1$.
Then, using standard spherical coordinates we have 
\begin{widetext}
\begin{eqnarray}
 \tilde{l}^a &=& (1, \frac{x}{r}, \frac{y}{r}, \frac{z}{r}) =(1, \sin(\theta) \cos(\phi), \sin(\theta) \sin(\phi), \cos(\theta) ),\\
 \tilde{n}^a &=& 
(1, -\sin(\theta) \cos(\phi), -\sin(\theta) \sin(\phi), -\cos(\theta) ) ,\\
 \tilde{m}^a &=& \eth_0 (\tilde{l}^a) =\frac{1}{\sqrt{2}} ( 0, -\cos(\theta) \cos(\phi)+ i\sin(\phi), -\cos(\theta) \sin(\phi)-i \cos(\phi), 
\sin(\theta) ),\\
 \bar{\tilde m}^a &= &\bar\eth_0 (\tilde{l}^a)=\frac{1}{\sqrt{2}} ( 0, -\cos(\theta) \cos(\phi)- i\sin(\phi), -\cos(\theta) \sin(\phi)+ i \cos(\phi), 
\sin(\theta) ). 
\end{eqnarray}
\end{widetext}
In these expressions we use the symbols $\eth_0$ and $\bar\eth_0$ to denote the edths operators
of the spheres of symmetry, with unit radius.

The transformation tetrad will be of the form
\begin{equation}
l^a = c_{l\tilde{n}}\tilde{l}^a-c_{l\bar{\tilde{m}}}\tilde{m}^a
-c_{l\tilde{m}}\bar{\tilde{m}}^a+c_{l\tilde{l}}\tilde{n}^a,
\end{equation}
\begin{equation}
n^a = c_{n\tilde{n}}\tilde{l}^a-c_{n\bar{\tilde{m}}}\tilde{m}^a 
-c_{n\tilde{m}}\bar{\tilde{m}}^a+c_{n\tilde{l}}\tilde{n}^a,
\end{equation}
\begin{equation}
m^a = c_{m\tilde{n}}\tilde{l}^a-c_{m\bar{\tilde{m}}}\tilde{m}^a
-c_{m\tilde{m}}\bar{\tilde{m}}^a+c_{m\tilde{l}}\tilde{n}^a, 
\end{equation}
\begin{equation}
\bar{m}^a = c_{\bar{m}\tilde{n}}\tilde{l}^a-c_{\bar{m}\bar{\tilde{m}}}\tilde{m}^a
-c_{\bar{m}\tilde{m}}\bar{\tilde{m}}^a+c_{\bar{m}\tilde{l}}\tilde{n}^a; 
\end{equation}
where the notation is $c_{l\tilde{n}}=l^a\tilde{n}_a$, and so on.
From these relations we can construct the transformation of the Ricci and Weyl scalars, 
but it is more convenient and easy to work with the spinor diad associated to the tetrad. 
Then the transformation of the dyads will be
\begin{equation}
\begin{split}
o^A=&A \tilde{o}^A+B \tilde{\iota}^A,\\
\iota^A=&C \tilde{o}^A+D \tilde{\iota}^A,
\end{split}
\end{equation}
together to the condition that ${o^A,\iota^A}$ conform a spinorial base, i.e.,
\begin{equation}\label{detunitario}
AD-BC=1,
\end{equation}

From the relations given in Appendix A, we get
\begin{equation}\label{eq:ABCD}
\begin{split}
A=&\frac{1}{\sqrt{2}}\sqrt{1-\sin(\theta)\sin(\phi)}e^{i\frac{\eta+\eta'+\pi}{2}} ,\\
B=&\sqrt{1+\sin(\theta)\sin(\phi)}e^{i\frac{\eta'-\eta+\pi}{2}},\\
C=&\frac{1}{2}\sqrt{1+\sin(\theta)\sin(\phi)}e^{i\frac{\eta-\eta'+\pi}{2}},\\
D=&\frac{1}{\sqrt{2}}\sqrt{1-\sin(\theta)\sin(\phi)}  e^{-i\frac{\eta'+\eta+\pi}{2}};\\
\end{split}
\end{equation}
where $\eta$ and $\eta'$ satisfies
\begin{eqnarray}
e^{i\eta}&=&\frac{-\cos(\theta)\sin(\phi)+i\cos(\phi)}{\sqrt{1-\sin^2(\theta)\sin^2(\phi)}},\\
e^{i\eta'}&=&\frac{\cos(\theta)+i\sin(\theta)\cos(\phi)}{\sqrt{1-\sin^2(\theta)\sin^2(\phi)}}\\
&=&\frac{z+ix}{J}=\cos(\vartheta)+i\sin(\vartheta), \nonumber
\end{eqnarray}
and in the last equality it was used the fact that 
$J=r\sqrt{1-\sin^2(\theta)\sin^2(\phi)}$,
(see Appendix A).
Note then that, $\eta'=\vartheta$.

The general transformation between tetrads induces the following transformation on the 
curvature scalar $\Phi_{00}$ and $\Psi_0$,
\begin{equation}
\begin{split}
 \Phi_{00}=&\Phi_{ABA'B'}o^Ao^Bo^{A'}o^{B'}\\ 
=& A^2\bar{A}^2\tilde\Phi_{00}
  + 2A^2\bar{A}\bar{B}\tilde\Phi_{01}
  + A^2\bar{B}^2\tilde\Phi_{02}\\
  &+ 2A\bar{A}^2B \tilde\Phi_{10}
  + 4A\bar{A}B\bar{B} \tilde\Phi_{11} 
  + 2AB\bar{B}^2\tilde\Phi_{12}\\
  &+ \bar{A}^2B^2 \tilde\Phi_{20}
  + 2B^2\bar{B}\bar{A} \tilde\Phi_{21}
  + B^2\bar{B}^2\tilde\Phi_{22},
\end{split}
\end{equation}
\begin{equation}
\begin{split}
\Psi_0=&\Psi_{ABCD}o^Ao^Bo^{C}o^{D}\\ 
=&A^4\tilde{\Psi}_0+4A^3\tilde{\Psi}_1+6A^2B^2\tilde{\Psi}_2\\
&+4AB^3\tilde{\Psi}_3+B^4\tilde{\Psi}_4 .
\end{split}
\end{equation}
In the spherically symmetric case these transformations simplify considerably and
finally, at linear order one has $\tilde\Phi_{22} = \frac{1}{4}\tilde\Phi_{00}$
so that
\begin{equation}\label{eq:scalarstab}
\begin{split}
\Psi_0= 3\frac{J^2}{r^2}\tilde{\Psi}_2(r)e^{2i\vartheta},
\end{split}
\end{equation}
\begin{equation}
\begin{split}
\Phi_{00}= \frac{2 J^2}{r^2}(\tilde\Phi_{11}-\frac{1}{4}\tilde\Phi_{00}) + \tilde\Phi_{00}.
\end{split}
\end{equation}

\subsubsection{The deflection angle in terms of spherically symmetric components of the curvature }
From (\ref{eq:simplealpha}), the function $\alpha(J)$ expressed in terms of the spherically symmetric null tetrad reads,
\begin{equation}\label{eq:lense-ext2}
\begin{split}
 \alpha(J)
= J \int_{-d_l}^{d_{ls}}  
 &\left[- \frac{3 J^2}{r^2}\tilde{\Psi}_2+\frac{2 J^2}{r^2}(\tilde{\Phi}_{11}
- \frac{1}{4}\tilde\Phi_{00}) \right. \\
&\; + \left. \tilde\Phi_{00}  \right] dy 
.
\end{split}
\end{equation}
Note that in this case, the integration is on the coordinate $y$, instead of
using arbitrary affine parameter. Also note that $r=\sqrt{J^2 + y^2}$.

This constitutes an important explicit relation for the bending angle
in terms of the curvature as seen in an spherically symmetric frame;
which is the natural frame for the sources of the gravitational lens.

\subsubsection{Expressions for the bending angle  in terms of energy-momentum components
and $M(r)$ }
Using eqs. (\ref{eq:phis}), (\ref{eq:phis11}), (\ref{eq:psi2rhoP}) and  (\ref{eq:lense-ext2})
we get an expression for 
the bending angle in terms of the mass, energy density, and spacelike components of the
energy-momentum tensor, namely
\begin{equation}\label{eq:lense-ext2-b}
\begin{split}
\alpha(J) = J \int_{-d_l}^{d_{ls}} 
 &\left[
 \frac{3 J^2}{r^2} \left( \frac{M(r)}{r^3} -  \frac{4 \pi}{3} \varrho(r) \right)
\right. \\
&\; \left. + 4 \pi \left(\varrho(r) +  P_r(r) \right)
\right] dy 
.\end{split}
\end{equation}

This is a new and useful relation for the deflection angle in terms of the
physical fields which are the sources of the gravitational lens.
It is also worth mentioning that this expression for the bending angle can also 
be deduced from the geodesic 
equation using standard techniques, as it is shown in Appendix B.

It is curious that the bending angle does not depend explicitly
on the tangential spacelike components of the energy-momentum tensor.

\subsubsection{The optical scalars in terms of spherically symmetric components of the curvature }
From equation (\ref{eq:lensscalarthin})  and (\ref{eq:lensscalarthin-g})
one obtains
\begin{equation}\label{eq:lensscalarthin-esf}
\kappa(J) = \frac{d_{ls} d_l}{d_s}
\int_{-d_l}^{d_{ls}} 
\left[
\frac{2 J^2}{r^2}(\tilde\Phi_{11}-\frac{1}{4}\tilde\Phi_{00}) + \tilde\Phi_{00}
\right]
dy
,
\end{equation}
and
\begin{equation}\label{eq:lensscalarthin-g-esf}
\gamma(J) =  -\frac{d_{ls} d_l}{d_s}
\int_{-d_l}^{d_{ls}} 
\left[
3\frac{J^2}{r^2}\tilde{\Psi}_2(r) 
\right]
dy
.
\end{equation}

These expressions give the optical scalars in terms of gauge invariant expressions for the
curvature components adapted to the symmetry of the matter distribution which is the
source of the gravitational lens.

\subsubsection{Expressions for the lens scalars  in terms of energy-momentum components
and $M(r)$}

In a similar way, the lens scalars, in terms of the spherically symmetric physical
fields, are given by
\begin{widetext}
\begin{equation}
\begin{split}
\kappa=&\frac{4\pi d_ld_{ls}}{d_s}
\int_{-d_l}^{d_{ls}} 
\left [\rho + P_r + \frac{J^2}{r^2}(P_t-P_r) \right]dy \\
\gamma = & \frac{ d_ld_{ls}}{d_s}
\int_{-d_l}^{d_{ls}} 
\frac{J^2}{r^2}\left [\frac{3 M}{r^3} - 4\pi(\rho+P_t-P_r)\right]dy
.
\end{split}
\end{equation}
\end{widetext}
These new expressions let us see explicitly the contributions of different components of the
energy-momentum tensor on the optical scalars.
One can see that a couple of terms disappear in the isotropic case in which $P_r=P_t$.

Our expressions are valid for generic energy-momentum distributions, 
but it is usually not considered in the literature the possible implications of non vanishing 
spacelike components of $T_{ab}$. In future works, we will consider the implications of 
models with non-trivial energy-momentum tensors on observed gravitational lenses.

\subsection{Two simple examples}

In order to show the application of our treatment of gravitational lens,
we will consider next two standard models that are often used in representing the source of 
gravitational lenses.
\subsubsection{A monopole mass (Schwarzschild)}
As a simple example let there be a monopole distribution characterized by a mass $M$, therefore a simple 
computation gives $\tilde{\Phi}_{00}=0$, and $\tilde{\Psi}_{2}=-\frac{M}{r^3}$,  then by 
considering that the observer and the source are far away, one can replace in the extremes 
of the integration (as is usually made) $d_s\rightarrow \infty$ and $d_l\rightarrow \infty,$ 
then 
\begin{eqnarray}
\hat{\Phi}_{00}&=&0,\\
\hat\psi_{0}&=&-2\int^{\infty}_{0}\frac{3J^2}{r^2} \tilde\Psi_2dy=\frac{4M}{J^2}, 
\end{eqnarray}
and by replacing into
eqs.(\ref{eq:lensscalarthin}), (\ref{eq:lensscalarthin-g}) and (\ref{eq:simplealpha}), 
we readily obtain the well known 
results
\begin{eqnarray}
\alpha(J)&=&\frac{4M}{J},\\
\kappa&=&0,\\
\gamma &=& 
\frac{d_ld_{ls}}{d_s} \frac{4M}{J^2} .
\end{eqnarray}
\subsubsection{The isothermal profile}
One simple model of dark matter that is used to explain the rotation curves of galaxies 
is the isothermal profile, which is defined by the density 
function
\begin{equation}
\rho=\frac{v^2_c}{4\pi r^2},
\end{equation}
where $v_c$ is the circular velocity.

Since $v_c\ll c$, the pressures in this model are negligible. 
Then we obtain,
\begin{eqnarray}
\hat\Phi_{00}&=&\int^{\infty}_{-\infty} \frac{v^2_c}{r^2}dy=\frac{v^2_c\pi}{J},\\
\hat\psi_0&=& 
 \int^{\infty}_{-\infty}\frac{2 J^2 v^2_c}{r^4}dy
=\frac{v^2_c\pi}{J}.
\end{eqnarray}
From these relations follow the well know results,
\begin{eqnarray}
\alpha&=&2\pi v^2_c,\\
\kappa&=&\frac{d_ld_{ls}}{d_s}\frac{v^2_c\pi}{J},\\
\gamma &=& 
\frac{d_ld_{ls}}{d_s}\frac{\pi v^2_c}{J}  .
\end{eqnarray}

\section{Final comments}\label{sec:final}

Several works on gravitational lensing reach up to the expressions that relate the optical scalars
with the curvature components in terms of the tetrad adapted to the motion of the photons;
we have here also presented expressions for the bending angle in terms of the curvature components.
Furthermore, we have presented above expressions for the optical scalars and deflection angle directly in terms
of the matter components of the sources of the gravitational lens, valid for an extended class
of matter distributions. In order to do that one has
to assume some structure for the source, so that in this first work on the subject, we have
treated the first natural model of spherical symmetry for the sources.
But in section \ref{sec:axisymm} we have presented expressions that are valid also for spheroidal
distributions, since we only required axissymmetry along the line of sight.

Our expressions circumvent several deficiencies as are: gauge dependence, lack of explicit expressions,
neglect of spacelike components of the energy-momentum tensor, etc.
It is probably worthwhile to remark that since the function $M(r)$ is determined in terms
of the $\varrho(r)$ by equation (\ref{eq:rho}), all our expressions are
explicit expressions in terms of the energy-momentum components of the matter
generating the gravitational lens.
As a trivial check of our equations we have presented two simple examples for which
the optical scalars and deflection angle are readily obtained.

The extension of this study to sources with different structure and to the cosmological background 
will be presented elsewhere.

\appendix
\section{}
\subsection{Tetrad transformations}
First, we note that
\begin{equation}
\frac{J^2}{r^2}=\frac{J^2}{J^2+y^2}=\frac{J^2}{J^2+r^2\sin(\theta)^2\sin(\phi)^2}
,\end{equation}
and solving this for $J^2/r^2$ we find
\begin{equation}
\frac{J^2}{r^2}=1-\sin(\theta)^2\sin(\phi)^2,
\end{equation}
The complete equations that satisfy the spinorial components are given by
\begin{equation}\label{eq:diadsys}
A\bar{A} = c_{l\tilde{n}}=\frac{1}{2}\left(-1+\sin(\theta)\sin(\phi)\right),
\end{equation}
\begin{equation}
 A\bar{B} = -c_{l\bar{{\tilde{m}}}}=-\frac{1}{\sqrt{2}}\left(\cos(\theta)\sin(\phi)-i\cos(\phi)\right),
\end{equation}
\begin{equation}
 B\bar{B} = c_{l\tilde{l}}=\left(-1-\sin(\theta)\sin(\phi)\right),
\end{equation}
\begin{equation}
 C\bar{C} = c_{n\tilde{n}}=\frac{1}{4}\left(-1-\sin(\theta)\sin(\phi)\right),
\end{equation}
\begin{equation}
 C\bar{D} = -c_{n\bar{\tilde{m}}}=\frac{1}{2\sqrt{2}}\left(\cos(\theta)\sin(\phi)-i\cos(\phi)\right),
\end{equation}
\begin{equation}
 A\bar{C} = c_{m\tilde{n}}=-\frac{1}{2\sqrt{2}}\left(-\cos(\theta)-i\sin(\theta)\cos(\phi)\right),
\end{equation}
\begin{equation}
 A\bar{D} = -c_{m\bar{\tilde{m}}}=\frac{1}{2}\left(\sin(\theta)+\sin(\phi)-i\cos(\theta)\cos(\phi)\right),
\end{equation}
\begin{equation}
 B\bar{C} = -c_{m\tilde{m}}=\frac{1}{2}\left(\sin(\theta)-\sin(\phi)-i\cos(\theta)\cos(\phi)\right),
\end{equation}
\begin{equation}
 B\bar{D} = c_{m\tilde{l}}=-\frac{1}{\sqrt{2}}(\cos(\theta)+i\sin(\theta)\cos(\phi)),
\end{equation}
\begin{equation}
 D\bar{D} = c_{n\tilde{l}}=\frac{1}{2}\left(-1+\sin(\theta)\sin(\phi)\right);
\end{equation}
and its complex conjugates together to the condition
\begin{equation}\label{eq:simplect}
AD-BC=1.
\end{equation} 

\section{}
\subsection{Deflection angle in terms of $T_{ab}$ from geodesic equation}

The four velocity vector of the particle has modulus
\begin{equation}\label{eq:uu}
 e^{2 \Phi} (\frac{dt}{d\lambda})^2
- \frac{1}{1 - \frac{2 M }{r}} (\frac{dr}{d\lambda})^2
- r^2 (\frac{d\varphi}{d\lambda})^2 = \kappa 
;
\end{equation}
where $\lambda$ is an affine parameter of the geodesic,
and we have already made use of the symmetry that allows
us to study just the motion in the equatorial plane $\theta = \frac{\pi}{2}$.
The constant $\kappa$ has values 1 for massive particles
and 0 for massless particles. This choice for $\kappa$ sets the unit
for the affine parameter for the massive particle case; however
the unit for the massless case remains undetermined.

There are also two integrals of motion.
$J$ is a constant of motion associated to the existence of a
rotational Killing vector which can be expressed by
\begin{equation}\label{eq:filam}
 J = r^2 \frac{d\varphi}{d\lambda};
\end{equation}
$E$ is another constant of motion associated to the existence of a
timelike Killing vector, which can be expressed by
\begin{equation}
 E = e^{2 \Phi} \frac{dt}{d\lambda} .
\end{equation}

Then equation (\ref{eq:uu}) takes the form
\begin{equation}\label{eq:uu2}
 e^{-2 \Phi} E^2
- \frac{1}{1 - \frac{2 M }{r}} (\frac{dr}{d\lambda})^2
-  \frac{J^2}{r^2} = \kappa 
;
\end{equation}
or
\begin{equation}\label{eq:uu3}
\begin{split}
&(\frac{dr}{d\lambda})^2
+  \left( \frac{J^2}{r^2}  - e^{-2 \Phi} E^2\right) (1 - \frac{2 M }{r}) \\ 
=&-\kappa (1 - \frac{2 M }{r})
;
\end{split}
\end{equation}
which can also be expressed as:
\begin{equation}\label{eq:uu4}
\begin{split}
&(\frac{dr}{d\lambda})^2
+
\frac{J^2}{r^2}
- \frac{J^2}{r^2} \frac{2 M }{r}
- \kappa\frac{2 M }{r}\\
&-  E^2 e^{-2 \Phi}  (1 - \frac{2 M }{r}) = 
-\kappa
 .
 \end{split}
\end{equation}

It is observed that the choice of the affine parameter $\lambda$ is related to the
definitions of the constants of motion $J$ and $E$.
Since $\Phi$ tends to zero in the asymptotic region, it is natural to take  $\lambda$ 
so that $E= 1$. This is equivalent to say that in the asymptotic region one has
$dt=d\lambda$. 

In this way there is no more freedom in the choice of units for $J$.
For an incident photon traveling in the $-y$ direction, with coordinate $x=x_0$,
the Newtonian expression for the angular momentum, for a unit mass particle gives
$J = r v \sin(\varphi+\frac{\pi}{2}) = r \cos(\varphi) = x_0$;
that is with this choice of affine parameter, $J$ has the meaning of asymptotic impact parameter
 $x_0$.

For convenience in the algebraic manipulation, let us define
\begin{equation}
 a_1(\Phi) \equiv 1 - e^{-2 \Phi} ;
\end{equation}
so that $e^{-2 \Phi} = 1- a_1(\Phi)$ in the above equation.

Therefore, for a photon, one can express (\ref{eq:uu4}) by
\begin{equation}\label{eq:uu5}
(\frac{dr}{d\lambda})^2
+
\frac{J^2}{r^2}
- J^2 \frac{2 M }{r^3}
+ \frac{2 M }{r} +\left(1- \frac{2 M }{r}\right) a_1(\Phi)
 = 1
.
\end{equation}

The corresponding potential for the motion of a photon is
\begin{equation}
 V_\ell = - J^2 \frac{ M }{r^3}
+ \frac{a_1(\Phi)}{2} + \frac{ M }{r} - \frac{ M }{r} a_1(\Phi)
;
\end{equation}
which, we remark, is an exact expression.

If one considers only linear departures from the flat metric, one
whould replace $e^{-2 \Phi} \approx (1 - 2 \Phi)$; and so one would obtain
\begin{equation}\label{eq:rlam}
 \begin{split}
 &\left( \frac{dr}{d\lambda}\right)^2 +
 \frac{J^2}{r^2}\\
+&2 \left[ 
 - \frac{\kappa M}{r} - \frac{J^2 M}{r^3} + E^2( \frac{M}{r}  + \Phi)
\right]= E^2 - \kappa 
.
\end{split}
\end{equation}

For the massless  case, by choosing the parametrization $\lambda$
so that $\frac{dt}{d\lambda}\rightarrow 1$ for $r\rightarrow \infty$
one has $E=1$, and therefore, one defines
\begin{equation}\label{eq:vlgener}
 V_\ell(r) = 
- \frac{J^2 M}{r^3} +  \frac{M}{r}  + \Phi
.
\end{equation}

The motion of a photon can
be deduced from the Lagrangian
\begin{equation}
\begin{split}
 L =& \frac{1}{2}  
\left( (\frac{dr}{d\lambda})^2 + r^2 (\frac{d\varphi}{d\lambda})^2 \right) 
- V_\ell(r) \\
=&
\frac{1}{2}
\left( (\frac{dx}{d\lambda})^2 + (\frac{dy}{d\lambda})^2 \right)
- V_\ell(r) 
;
\end{split}
\end{equation}
where in the last equality we have used Cartesian like coordinate
system with $r=\sqrt{x^2 + y^2}$.
This system obviously has the integral of motion (\ref{eq:filam})
with Lagrangian energy $\mathcal{E} = \frac{E^2}{2}$.

The equations of motion are:
\begin{equation}
 \frac{d v_x}{d\lambda} =
- \frac{x}{r} \frac{dV_\ell}{dr} ,
\end{equation}
\begin{equation}
 \frac{d v_y}{d\lambda} =
- \frac{y}{r} \frac{dV_\ell}{dr}
;
\end{equation}
with the velocity notation $v_x = \frac{d x}{d\lambda}$
an d$v_y = \frac{d y}{d\lambda}$.

Let us assume the initial conditions:
$x= x_0$, $y >> 2 M$, $v_x=0$ and $v_y = -1$.
Then, in this case, $x_0$ is the impact parameter, so that $J=x_0$.

After passing through the gravitational lens, the trajectory
will be deflected so that, in the asymptotic region one would have
$v_x|_\infty =  \delta v$ and $v_y|_\infty =  -\sqrt{1 -(\delta v)^2 }$;
since the photon must travel at the velocity of light.
Then the bending angle can be calculated from
\begin{equation}
 \alpha = -\arctan \frac{v_x|_\infty}{v_y|_\infty}
\approx -\frac{v_x|_\infty}{v_y|_\infty}
\approx -\delta v
.
\end{equation}

The variation in the velocity can be calculated from
\begin{equation}
 \delta v =  -\int_{\lambda_o}^{\lambda_f}  \frac{x}{r}   \frac{dV_\ell}{dr} d\lambda
=
 - \int_{-d_l}^{d_{ls}}  \frac{x}{r}   \frac{dV_\ell}{dr} dy
;
\end{equation}
where we have taken $d\lambda=dy$.

The coordinate system has origin at the center of the spherical symmetry.
In the approximation of a lens contained in a plane, the center is in this plane.

To consider the equation of motion of a massless particle in the more general
case, we also use the equations of motion in the Cartesian like
coordinate system, where now the potential is given by (\ref{eq:vlgener});
so that
\begin{equation}
\begin{split}
 \frac{d V_\ell}{dr} =&
2 \frac{J^2 }{r^3} \frac{M}{r} 
+(1 - \frac{J^2 }{r^2}) \left( \frac{\frac{dM}{dr}}{r} - \frac{M}{r^2}\right)
+ \frac{d\Phi}{dr} \\
=&
3 \frac{J^2 M}{r^4} - \frac{M}{r^2}
+(1 - \frac{J^2 }{r^2}) 4 \pi r \rho\\
&+\frac{M + 4\pi r^3 P_r}{r^2(1 - \frac{2 M}{r})}
;
\end{split}
\end{equation}
and in the linear regime one has
\begin{equation}\label{eq:force1}
\begin{split}
 - \frac{d V_\ell}{dr}
=&
\frac{3 J^2}{r} \left(- \frac{M(r)}{r^3}
+  \frac{4 \pi}{3} \rho(r) \right)\\
&- 4 \pi r  \left(\rho(r) +  P_r(r) \right)
.
\end{split}
\end{equation}

Finally the deflection angle is given by
\begin{widetext}
\begin{equation}\label{eq:lense-ext2-ap}
\begin{split}
 \alpha =& -\delta v =
 \int_{-d_l}^{d_{ls}}  \frac{x_0}{r} \frac{d V_\ell}{dr} dy \\
=&  \int_{-d_l}^{d_{ls}}  \frac{x_0}{r} 
 \left[
- \frac{3 J^2}{r} \left(- \frac{M(r)}{r^3} +  \frac{4 \pi}{3} \rho(r) \right)
+4 \pi r  \left(\rho(r) +  P_r(r) \right)
\right] dy 
;
\end{split}
\end{equation}
\end{widetext}
where it is understood that $r=\sqrt{x_0^2 + y^2}$.

This coincide with expression (\ref{eq:lense-ext2-b}) appearing above.

\section*{Acknowledgments}

We acknowledge support from CONICET and SeCyT-UNC.


\end{document}